\documentclass[12pt]{article}  
\usepackage{a4}
\usepackage{graphicx}
\usepackage{subfigure}
\usepackage[centertags]{amsmath}
\usepackage{amssymb}

\newcommand{\barre}[1]{%  
        \setbox1=\hbox{$#1$} \dimen2=\ht1 \dimen3=\dp1 \dimen4=\wd1  
        \setbox2=\hbox{\sl /}  
        \dimen1=\wd1 \advance\dimen1 by -\wd2 \divide\dimen1 by 2  
        \advance\dimen1 by \wd2 \advance\dimen1 by 0.4pt  
        \setbox3=\hbox to \wd1{\hss \box1 \kern -\dimen1 \box2\hss}  
        \ht3=\dimen2 \dp3=\dimen3 \wd3=\dimen4  
        \box3  
        }  
\begin{document}  
\begin{flushright}  
                     hep-th/0112193
\end{flushright}  
\vskip 2cm    

\begin{center}  
{\huge D-branes on a Deformation of $SU\! \left(
  2 \right) $}   
\vspace*{5mm} \vspace*{1cm}   
\end{center}  
\vspace*{5mm} \noindent  
\vskip 0.5cm  
\centerline{\bf Stefan F\"orste}
\vskip 1cm
\centerline{\em Physikalisches Institut, Universit\"at Bonn}
\centerline{\em Nussallee 12, D-53115 Bonn, Germany}
\vskip2cm
  
\centerline{\bf Abstract}  
\vskip .3cm
We discuss D-branes on a line of conformal field theories connected by
an exact marginal deformation. The line contains an $SU\!\left(
  2\right)$ WZW model and two mutually T-dual $SU\!\left(
  2\right)/U\!\left( 1\right)$ cosets times a free boson. We find the
D-branes preserving 
a $U\!\left( 1\right)$ isometry, an $F$-flux quantization condition
and conformal invariance. Away from the $SU\!\left( 2\right)$ point
a $U\!\left( 1\right)\times U\!\left( 1\right)$ symmetry is broken to
$U\!\left( 1\right) \times {\mathbb Z}_k$, i.e.\ continuous rotations
of branes are accompanied by rotations along the branes. Requiring
decoupling of the cosets from the free boson at the endpoints of the
deformation breaks 
the continuous rotation of branes to ${\mathbb Z}_k$. At the
$SU\!\left( 2\right)$ point the full $U\!\left( 1\right)\times
U\!\left( 1\right)$ symmetry is restored.
This
suggests the occurrence of phase transitions for branes at angles in
the coset model, at a semiclassical level. We also discuss briefly
the orientifold planes along the deformation line.

\vskip .3cm

%%%%%%%%%%%%%%%%%%%%%%%%%%%%%%%%%%%%%%%%%%%%%%%%%%%%%%%%%%%%%%%%%%%%%%%%%%%%%  
  
\newpage  

\section{Introduction}

In the present article we consider D-branes on a family of conformal
field theories connecting an $SU\!\left( 2\right)$ WZW model with two
$SU\!\left( 2\right)/U\!\left( 1\right)$ coset models differing by the
way the $U\!\left( 1\right)$ is embedded into $SU\!\left( 2\right)$. 
For the undeformed WZW model and the coset models possible D-brane
setups have been discussed before. Geometrically the $SU\!\left(
  2\right)$ is a three dimensional sphere $S^3$. Hence, the isometries
form an $SO\!\left( 4\right)$ group. Wrapping D2 branes on two
dimensional spheres in
$S^3$ one can set up D-brane configurations preserving an $SO\!\left(
  3\right)$ subgroup of the isometry transformations. The
corresponding boundary conditions where given
in\cite{Kato:1996nu}\footnote{For a `pre-brane-era' discussion of open
  strings on the $SU\!\left( 2\right)$ manifold see\cite{Behrndt:1993fh}.}
but
not interpreted as belonging to D2-branes. This point was clarified
in\cite{Alekseev:1998mc} where the general solutions to the boundary
conditions were given. It was also pointed out that the position of
the D2 brane can take only discrete values due to a topological
argument. 

Because a two dimensional sphere is contractable on a three
dimensional sphere, there is a potential stability problem with the
above discussed D2 branes. This puzzle was solved in
\cite{Bachas:2000ik,Pawelczyk:2000ah} and it was pointed out that the
quantization of the D2-brane position corresponds to an $F$-flux
quantization condition, where $F$ is the fieldstrength of a
$U\!\left(1\right)$ gauge field living on the brane.   

The $S^3$ geometry can be deformed to a class of geometries with
$U\!\left( 1\right)\times U\!\left( 1\right)$ isometry. In the
conformal field theory this corresponds to an exact marginal
deformation \cite{Hassan:1992gi},\cite{Giveon:1993ph}. Therefore, a
natural way to deform the D2-branes on $S^3$ is given by the
prescription that they should preserve a $U\!\left( 1\right)$
isometry. By imposing this condition we will find deformed branes
satisfying an $F$-flux quantization condition and preserving conformal
invariance. 

At the end of the deformation line the geometry degenerates into an
$SU\!\left( 2\right)/U\!\left( 1\right)$ coset times a decoupled
$U\!\left( 1\right)$ of vanishing size. D-branes on the cosets were
discussed in\cite{Maldacena:2001ky}, \cite{Gawedzki:2001ye},
\cite{Parvizi:2001xe}, \cite{Elitzur:2001qd},
\cite{Fredenhagen:2001kw}, \cite{Ishikawa:2001zu},
\cite{Kubota:2001ai}. An important 
information is that the $U\!\left( 1\right)$ isometry of
rotations in the coset geometry is broken to ${\mathbb Z}_k$, because
the corresponding modulus couples as a `theta angle' in the gauged WZW
model\cite{Maldacena:2001ky}. This stabilizes D-branes which would be
unstable if continuous rotations were allowed. Observing that the line
of marginally deformed models can be viewed as an $SU\!\left( 2\right)
\times U\!\left( 1\right) / U\!\left( 1\right)$
model\cite{Sfetsos:1993ka}, we find that a $U\!\left( 1\right)\times
U\!\left( 1\right)$ 
isometry is broken to $U\!\left( 1\right)\times{\mathbb Z}_k$ along
the line of deformed 
models except for the point corresponding to the undeformed case. The
reason is that for
the undeformed case the size of the extra $U\!\left( 1\right)$ becomes
a modulus and continuous shifts in the `theta' angle can be absorbed
in a rescaling of the gauge group size. Hence, phase transitions can
be triggered by rotating branes in the undeformed model.

The paper is organized as follows. In section 2 we review the
$SU\!\left( 2\right)$ preserving D-branes in the WZW model. This is
done on an explicit level. We also discuss the $F$-flux quantization
condition. Section 3 briefly recalls the family of conformal field
theories obtained by exact marginal deformations of the closed string
on $SU\!\left( 2\right)$. In the fourth section we introduce D-branes
corresponding to the deformed $SU\!\left( 2\right)$ preserving D2
branes. We check that conformal invariance and the $F$-flux
quantization condition are preserved. Section 5 is devoted to an
investigation of the moduli spaces along the deformation line. By
imposing decoupling of the free boson at the ends of the deformation
line, we argue that at one end of the line branes can be 
rotated only by discrete angles whereas at the other end of the
deformation line Wilson lines can be shifted only by discrete
values. At the $SU\!\left( 2\right)$ point those moduli become
continuous.  
In section 6 we discuss the A- and B-branes
of\cite{Maldacena:2001ky} along the deformation line and suggest that
phase transitions might by associated with the rotation of branes at
the $SU\!\left( 2\right)$ point. In a seventh section the possible
orientifold fixed planes are briefly described. We summarize our
results in a concluding section 8.

\section{Recap of D-branes on an $SU(2)$ WZW model - coordinate
  dependent description}\label{recap} 

Let us first discuss the WZW model for worldsheets without
boundaries. The action is given by
\begin{equation}
S = S^{kin} + S^{WZ} = \frac{k}{2\pi}\left[ \int_{\Sigma} d^2 z {\cal
    L}^{kin} + \int_{\cal B} \omega^{WZ} \right] ,
\end{equation}
where $k$ is the level of the WZW model. The worldsheet $\Sigma$ is
parameterized by  
\begin{equation}
z_{\pm} = \frac{\tau \pm \sigma}{2}
\end{equation}
and ${\cal B}$ is a three dimensional manifold whose boundary is 
$\Sigma$. The requirement of the theory to be independent of the
particular choice of ${\cal B}$ leads to the quantization condition
that the 
level $k$ should be integer valued\cite{Witten:ar}. The Lagrangian for
the kinetic term reads
\begin{equation}
{\cal L}^{kin} = tr \left( \partial_+ g \partial_- g^{-1}\right)
\end{equation}
whereas the integrand in the Wess Zumino term is given as
\begin{equation}
\omega^{WZ} = \frac{1}{3}tr\left[ \left( g^{-1} dg\right)^3\right] ,
\end{equation}
where the power of three is understood with respect to the wedge
product of differentials.  

We parameterize the $SU\! \left( 2\right)$ group element $g$ with
``spherical coordinates'', i.e.
\begin{equation}
g = \cos \chi + i \sin \chi \cos \vartheta\; \sigma^1 + i \sin \chi
\sin\vartheta \cos\varphi \; \sigma^2 + i \sin \chi \sin \vartheta \sin
\varphi \; \sigma^3 ,
\end{equation}
where $\sigma^i$ are the Pauli matrices and the first term is
understood to be multiplied by the identity matrix. The parameter
ranges are
\begin{equation}
\chi = 0 \ldots \pi \; ,\; \vartheta = 0\ldots \pi\; ,\, \varphi =
0\ldots 
2\pi .
\end{equation}
For later use let us give the following differential on the group
manifold 
\begin{eqnarray}
g^{-1} dg & = &\left\{ \cos \vartheta d\chi - \sin \chi \cos \chi \sin
    \vartheta d\vartheta + \sin^2 \chi \sin^2 \vartheta d\varphi \right\}i
    \sigma^1 \nonumber \\
& & +\left\{ \sin \vartheta \cos \varphi d\chi +\left( \sin \chi \cos
    \chi \cos \vartheta \cos \varphi - \sin^2 \chi \sin \varphi\right)
    d\vartheta \right. \nonumber\\
& & \; \left. -\left( \sin^2 \chi \sin \vartheta \cos\vartheta \cos
    \varphi +
    \sin\chi \cos\chi \sin\vartheta \sin \varphi\right)
    d\varphi\right\} i \sigma^2 \nonumber \\
& & + \left\{ \sin \vartheta \sin \varphi d\chi + \left( \sin \chi
    \cos \chi \cos \vartheta \sin \varphi + \sin^2 \chi  \cos \varphi
    \right) d\vartheta \right. \nonumber\\
& & \; \left. + \left( \sin \chi \cos \chi \sin \vartheta
    \cos \varphi - \sin^2 \chi \sin \vartheta \cos \vartheta \sin
    \varphi \right) d\varphi \right\} i\sigma^3   \label{rightKC}
\end{eqnarray}
The differential $dg g^{-1}$ can be obtained from (\ref{rightKC}) by
multiplication with $-1$ and replacing $\chi$ by $-\chi$. Hence, from
the expression (\ref{rightKC}) one can read off the explicit form of
the Kac-Moody currents
\begin{eqnarray}
J &=& -\partial_+ g g^{-1} \equiv J_\mu \partial_+ X^\mu ,\\
  \bar{J} & = & g^{-1}\partial_- g \equiv \bar{J}_\mu \partial_- X^\mu ,
\label{glucon}
\end{eqnarray}
where $X^\mu$ denotes the three targetspace coordinates $\chi,
\vartheta , \varphi $.

For the action of the closed string on the
$SU\!\left( 2\right)$ manifold we find
\begin{eqnarray}
S^{WZW} & = & \frac{k}{4\pi} \int d^2 \sigma \left\{ \partial_\alpha
  \chi 
  \partial^\alpha \chi + \sin^2 \chi \partial_\alpha \vartheta
  \partial^\alpha \vartheta +\sin^2 \chi \sin^2 \vartheta
  \partial_\alpha \varphi \partial^\alpha \varphi\right. \nonumber\\ 
& & \; \left. - 2 \left( \chi -\frac{\sin 2\chi}{2}\right)\sin
  \vartheta \left( \partial_\tau \vartheta \partial_\sigma\varphi
  -\partial_\sigma\vartheta \partial_\tau \varphi\right)\right\} .
\label{expliwzw}
\end{eqnarray}
The
worldsheet indices $\alpha ,\beta =0,1$ are raised and lowered with
the Minkowski metric $diag\left( 1, -1\right)$. Note, that we are free
to add terms to the Lagrangian which vanish 
upon integration over the compact worldsheet $\Sigma$, i.e.\ we can
add terms corresponding to an antisymmetric tensor field (a $B$-field)
which is pure gauge. Our $B$-field reads
\begin{equation}
B = k\alpha^\prime \left( \chi - \frac{\sin 2\chi}{2}\right) \sin
\vartheta d\vartheta \wedge d\varphi .
\label{schweigeB}
\end{equation}
This choice is taken from\cite{Bachas:2000ik} and motivated by the
observation that the Aharanov-Bohm phase of a fundamental string
wrapping around this potential is an integer multiple of $2\pi $ due
to the quantization condition on $k$.

After we have written the action of the WZW model in terms of
integrals over the worldsheet only, we can try to describe open strings
on $SU\!\left( 2 \right)$ via just replacing the closed worldsheet in
(\ref{expliwzw}) by a worldsheet with boundary (the upper half plane). 
Then we have to impose boundary conditions which do not spoil the
conformal invariance, i.e.\ the Dirac-Born-Infeld (DBI) action on the 
corresponding D-brane should be minimal\cite{Leigh:jq}.
We do not explore directly all possible boundary conditions but
require instead that out of the $SU\!\left( 2\right)\times SU\!\left(
  2\right)$ 
symmetry of the WZW model the boundary conditions preserve a residual
$SU\!\left( 2\right)$ symmetry. Afterwards we will check that
conformal invariance is preserved.

Let us now work out explicitly the possible boundary conditions
imposed by the requirement that they preserve a residual
$SU\!\left( 2\right)$ symmetry. The corresponding gluing conditions in
the closed string channel have been worked out
in\cite{Kato:1996nu}. In the open string channel this results in
the following boundary conditions for the Kac-Moody
currents\cite{Cardy:ir, Alekseev:1998mc,Stanciu:1999id},
\begin{equation}
\left( J_\mu - \bar{J}_\mu\right) \partial_\tau X^\mu + \left( J_\mu
  +\bar{J}_\mu \right) \partial_\sigma X^\mu = 0.
\end{equation}
Plugging in our explicit parameterization of the $SU\! \left( 2\right)$
elements leads to the three equations (corresponding to the factors in
front of the three Pauli matrices)
\begin{eqnarray}
0& = & -\cos\vartheta \,\partial_\tau \chi  +\sin\chi \cos \chi \sin 
 \vartheta\,\partial_\tau \vartheta +\sin^2 \chi \sin^2 \vartheta
 \,\partial_\sigma \varphi ,\label{bc1} \\
0 & = & -\sin\vartheta \cos\varphi \,\partial_\tau \chi - \sin\chi
 \cos\chi \cos\vartheta \cos \varphi \,\partial_\tau \vartheta + \sin
 \chi \cos\chi \sin \vartheta \sin \varphi\, \partial_\tau
 \varphi\nonumber\\
& & \;\; -
 \sin^2 \chi \sin \varphi\, \partial_\sigma \vartheta - \sin^2 \chi \sin
 \vartheta \cos \vartheta \cos \varphi \,\partial_\sigma \varphi , \\
0 & =& -\sin \vartheta \sin \varphi \, \partial_\tau \chi - \sin \chi
 \cos \chi 
 \cos \vartheta \sin \varphi\, \partial_\tau \vartheta - \cos \chi \sin
 \chi \sin \vartheta \cos \varphi\, \partial_\tau \varphi\nonumber
 \label{bc2}\\ 
& & \;\; + \sin^2 \chi
 \cos \varphi \, \partial_\sigma \vartheta - \sin^2 \chi \sin \vartheta
 \cos \vartheta \sin \varphi\, \partial_\sigma \varphi . \label{bc3}
\end{eqnarray}
We observe that the coordinate $\chi$ does not enter the boundary
conditions with its transverse derivative $\partial_\sigma
\chi$. Therefore, 
we choose Dirichlet boundary conditions fixing $\chi$ to some constant
value
\begin{equation}
\chi_{\left|\sigma =0\right.} = \chi_0 .
\end{equation}
The rest of the equations (\ref{bc1})--(\ref{bc3}) yields inhomogeneous
Neumann conditions for $\vartheta$ and $\varphi$
\begin{eqnarray}
\frac{1}{k\alpha^\prime}G_{\vartheta\vartheta}\, \partial_\sigma
  \vartheta & = & \frac{\sin 
  2\chi_0}{2} \sin \vartheta \, \partial_\tau \varphi \\ 
\frac{1}{k\alpha^\prime}G_{\varphi \varphi} \,\partial_\sigma \varphi
  & = &  -\frac{\sin 
  2\chi_0}{2} \sin\vartheta 
\, \partial_\tau \vartheta ,
\end{eqnarray}
where $G_{\varphi\varphi} = \sin^2 \vartheta G_{\vartheta\vartheta} =
k\alpha^\prime\sin^2 \chi_0 \sin^2 \vartheta$ are the metric components of the
targetspace. 
Note, that the resulting boundary value of the group element can be
written as
\begin{equation}
g\left( \tau \right) = k\left( \tau\right)   e^{i\chi_0 \sigma^3}
k^{-1}\left( \tau\right) ,
\label{conclass}
\end{equation}
with
\begin{equation}
k\left( \tau \right) = e^{i\left( \frac{\pi}{4} -\frac{\varphi\left(
        \tau\right) }{2}\right)\sigma^1} e^{i\left(
        \frac{\vartheta\left(\tau\right)}{2} -
        \frac{\pi}{4}\right)\sigma^2} .
\end{equation}
Thus, the statement that the D-brane is the conjugacy class of a fixed
group element\cite{Alekseev:1998mc} is verified explicitly. In
addition the expression (\ref{conclass}) allows us to read off the
allowed quantized values of $\chi_0$ which are known from CFT
analysis\cite{Alekseev:1998mc, Gawedzki:1999bq,
  Elitzur:2001qd}\footnote{At least in the semiclassical treatment,
  one obtains zero dimensional branes at $\chi =0,\pi$. Since the
  coordinates degenerate there, it is problematic to differ between
  $D0$ branes (Dirichlet conditions in all directions) and collapsed
  higher dimensional branes. A coordinate independent way to see this
  problem is given by the observation that the conjugacy classes of
  plus or minus the identity are plus or minus the identity. We will
  exclude the zero dimensional branes from our discussion, mostly.} 
\begin{equation}
\chi_0 \in \frac{\pi {\mathbb Z}}{k}.
\label{quantization}
\end{equation}
As it stands our boundary conditions cannot be derived from
(\ref{expliwzw}) (with the integration region being the upper half
plane) by the Hamiltonian principle. In order to achieve consistency we
have to add a boundary term such that the action for the open string
on the WZW model finally reads
\begin{eqnarray}
S^{WZW}_{open} & = & \frac{k}{4\pi} \int d^2 \sigma \left\{ \partial_\alpha
  \chi 
  \partial^\alpha \chi + \sin^2 \chi \partial_\alpha \vartheta
  \partial^\alpha \vartheta +\sin^2 \chi \sin^2 \vartheta
  \partial_\alpha \varphi \partial^\alpha \varphi\right. \nonumber\\ 
& & \; \left. - 2 \left( \chi-\chi_0 -\frac{\sin 2\chi}{2}\right)\sin
  \vartheta \left( \partial_\tau \vartheta  \partial_\sigma\varphi
  -\partial_\sigma \vartheta \partial_\tau \varphi\right)\right\} .
\label{wzw-flux}
\end{eqnarray}
The term with the explicit dependence on the position $\chi_0$ of the D-brane
can be written as an integral over the
real line. Thus, the additional term corresponds to an $F$-flux on the
worldvolume of the D-brane given by\cite{Bachas:2000ik}
\begin{equation}
F = - \frac{k}{2\pi}\chi_0 \sin \vartheta d\vartheta \wedge d\varphi .
\label{Fquant}
\end{equation}
Comparison with (\ref{quantization}) shows that the flux $F$ is
quantized. The flux quantization condition looks like a gauge ($B\to B
+d\Lambda$) dependent condition. However, gauges changing the $F$-flux
are not single valued and multi valued gauge
transformations shift the flux by an integer
amount\cite{Bachas:2000ik}. Moreover, it is easy to check that the DBI
action is at a stationary point and thus conformal invariance is
preserved\cite{Leigh:jq}. (We will present a more detailed discussion
of this point when introducing branes in the deformed models, later.)
%Below, we
%will formulate the model in different coordinates and a different
%gauge (connected to the present one by a non single valued
%transformation). Therefore, it is useful to write down a quantization
%condition which is not affected by (even multi valued) gauge
%transformation. Such a condition is satisfied by the energy of the D2
%branes\cite{Bachas:2000ik}, 
%
%\begin{equation}
%E = T^{(2)} \int_0 ^\pi d\vartheta \int_0 ^{2\pi} d\varphi \sqrt{
%  \det\left( \hat{G} + \hat{B} + 2\pi \alpha^\prime F\right)} =
%  T^{(2)} 4\pi\alpha^\prime \sin \chi_0 ,
%\label{Equant}
%\end{equation}
%
%with $\chi_0$ satisfying (\ref{quantization}). The hat on bulk fields
%denotes the field induced on the brane.  

In summary, we have reviewed the possible boundary conditions
preserving an $SU\!\left( 2\right)$ 
symmetry and satisfying a quantization condition of the $F$-flux on
the 
D-brane. This will give us some guide for finding possible
D-branes on a deformed $SU\!\left( 2\right)$ manifold later.

\section{The closed string on a deformed $SU\!\left( 2\right)$}

In this section we review the results of an exact marginal
deformation of the WZW model\cite{Hassan:1992gi}. We closely follow
the presentation in \cite{Giveon:1993ph}.
The exact marginal deformation is obtained by integrating an
infinitesimal perturbation with an operator of conformal dimension $(1
,1)$. The integration of the perturbation is much
easier in different coordinates. Therefore, we parameterize the group
element as follows
\begin{equation}
g = \cos x \cos \tilde{\theta} -i\sigma^1 \, \sin x \sin \theta +
i\sigma^2\, \sin x \cos \theta + i \sigma^3 \cos x \sin \tilde{\theta} ,
\label{depara}
\end{equation}
with the parameter ranges
\begin{equation}
x = 0\ldots \frac{\pi}{2}\; ,\; \theta = -\pi \ldots \pi \; ,\;
\tilde{\theta} = -\pi \ldots \pi .
\end{equation}
In this parameterization the action of the closed string on the group
manifold reads
\begin{eqnarray}
S^{WZW} & = & \frac{k}{2\pi} \int d^2z \left\{ \partial_+ x\partial_- x +
  \sin^2 x \partial_+ \theta \partial_- \theta + \cos^2 x \partial_+
  \tilde{\theta} \partial_- \tilde{\theta} \right. \nonumber \\
& & \left. \; \;  +\cos^2 x \left( \partial_+
  \theta \partial_- \tilde{\theta} - \partial_+
  \tilde{\theta}\partial_- \theta \right)\right\} .
\label{GiKiWZW}
\end{eqnarray}
The advantage of the above parameterization is that a chiral and an
anti-chiral current are manifest in (\ref{GiKiWZW}). These are
\begin{eqnarray}
J & = & k\left(  \sin ^2 x \partial_+ \theta - \cos^2 x \partial_+
  \tilde{\theta}\right) , \label{udchiral}\\
\bar{J} & = & k \left( \sin^2 x \partial_- \theta + \cos^2 x
  \partial_- \tilde{\theta}\right) \label{udanti}.
\end{eqnarray}
Hence, a good candidate for a marginal deformation is the product
$J\bar{J}$. Indeed, such a perturbation can be integrated to finite
deformations with the resulting action
\begin{eqnarray}
S^R & = & \frac{k}{2\pi} \int d^2z \left\{ \partial_+ x\partial_- x +
  \frac{\sin^2 x}{\cos^2 x + R^2 \sin^2 x} \partial_+ \theta
  \partial_- \theta \right. \nonumber \\
& & \left. \; \; + \frac{ R^2 \cos^2 x}{\cos^2 x + R^2 \sin^2 x}
  \partial_+  
  \tilde{\theta} \partial_- \tilde{\theta}   +\frac{\cos^2 x}{\cos^2 x
  + R^2 \sin^2 x} \left( 
  \partial_+ 
  \theta \partial_- \tilde{\theta} - \partial_+
  \tilde{\theta}\partial_- \theta \right)\right\} .
\label{GiKiDefo}
\end{eqnarray}
In addition a nontrivial dilaton $\Phi$ is generated by the deformation
according to
\begin{equation}
e^{-2\Phi\left( R=1\right)} \sqrt{G\left( R=1\right)} =
e^{-2\Phi\left( R\right)} \sqrt{ G\left( R\right)} ,
\label{dildefo}
\end{equation}
where $R=1$ corresponds to the undeformed case with a constant
dilaton, and $G$ is the determinant of the target space metric.

The deformation breaks the original $SU\!\left( 2\right) \times
SU\!\left( 2\right)$ symmetry to a $U\!\left( 1\right) \times
U\!\left( 1\right)$ symmetry. The corresponding chiral and anti-chiral
currents are manifest in the present parameterization and read
\begin{eqnarray}
J\left( R\right) & = & \frac{J\left( R =1\right)}{\cos^2 x + R^2
  \sin^2 x} ,\label{chiral}\\
\bar{J}\left( R \right) & = & \frac{\bar{J}\left( R=1\right)}{\cos^2 x +R^2
  \sin^2 x}
\label{antichiral}  .
\end{eqnarray}
That these currents are conserved follows from a combination of the
$\theta$ and $\tilde{\theta}$ equation of motion and hence is not
sensitive to the $x$ dependent dilaton.

\begin{figure}
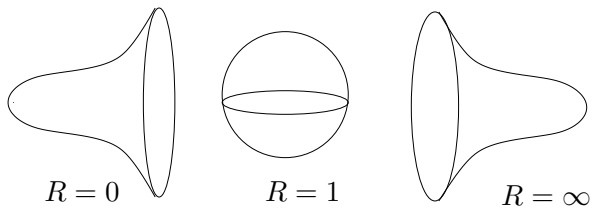
  
\begin{center}
\input pico.pstex_t
\end{center}
\caption{The ``middle'' and end points of the deformed model. At the
  endpoints there is an additional circle with vanishing radius.}
\label{fig:closeddefo}
\end{figure}

At the endpoints of the marginal deformation at $R=0$ and $R=\infty$
the geometry factorizes into a two dimensional manifold
times a circle (with vanishing radius). Hence, we obtain the picture
drawn in figure \ref{fig:closeddefo}.
%%%%%%%%%%%%%%%%%%%%%%%%%%%%%%%%%%%%
%%%%%%%%%%%%%%%%%%%%%%%%%%%%%%%%%%%% 
The geometries at the endpoints of the deformation are related by
T-duality with respect to $\theta$ ($\tilde{\theta}$) (and renaming
$\theta \leftrightarrow \tilde{\theta}$ afterwards).

To close this section we should note that the $B$ field in (\ref{GiKiWZW})
does not coincide with (\ref{schweigeB}) transformed to the
coordinates $\theta$, $\tilde{\theta}$, $x$. In these coordinates one
obtains a rather lengthy expression for (\ref{schweigeB}) and we did
not attempt to find the additional gauge transformation connecting
(\ref{schweigeB}) with the $B$ field in (\ref{GiKiWZW}). This gauge
transformation should not be single valued\cite{Bachas:2000ik}. Therefore,
we cannot expect to find the $F$-flux quantization condition
(\ref{Fquant}) without performing a multi valued gauge transformation.   

\section{The open string on a deformed $SU\!\left( 2\right)$}

In this section we would like to investigate the question whether and
where we can add D-branes to the deformed geometry discussed in the
previous section. As a guideline we take first the requirement that
the $U\!\left( 1\right) \times U\!\left( 1\right)$ symmetry of the
deformed model is broken by the boundary conditions to a residual
$U\!\left( 1\right) $ symmetry. This provides gluing conditions of the
chiral (\ref{chiral}) with the anti-chiral (\ref{antichiral}) currents.
We choose these gluing conditions such that in the undeformed case they
are solved by our previous boundary conditions. Performing a
coordinate transformation one can verify that a condition satisfying
our criteria is given by (\ref{glucon}) with $J ,\bar{J}$ as given in 
(\ref{chiral}), (\ref{antichiral}), (\ref{udchiral}),
(\ref{udanti}). Hence, the gluing conditions are independent of the
deformation parameter $R$. Expressed in the coordinates $x, \theta ,
\tilde{\theta}$ the gluing condition reads
\begin{equation}
-\cos^2x\, \partial_\tau \tilde{\theta} + \sin^2 x\, \partial_\sigma
\theta = 0.
\label{gludefo}
\end{equation}
As it stands, this is one condition for three coordinates. However,
our knowledge of the undeformed model leads us to impose in addition
the Dirichlet condition\footnote{Expressed in the coordinates of the
  previous section, $\chi$ is the only direction which does not enter
  (\ref{gludefo}) with its $\sigma$ derivative. The particular
  boundary value for $\chi$ could depend on the deformation parameter
  $R$. We will argue below that this is prohibited by the quantization
  condition on the $F$-flux.}
\begin{equation}
\cos \chi_0 = \cos x \cos\tilde{\theta} .
\label{Ddefo}
\end{equation}
Now, it is natural to read (\ref{gludefo}) as
inhomogeneous Neumann condition on $\theta$
\begin{eqnarray}
\frac{1}{k\alpha^\prime} G_{\theta\theta}\,\partial_\sigma \theta & =
& 
\frac{\cos^2 x}{\cos^2 x +R^2 \sin^2 x}\,\partial_\tau
\tilde{\theta}\nonumber \\
 & = & -\frac{\sin x \cos x\cos
  \chi_0}{\left(\cos^2 x +R^2 \sin^2 x\right)\sqrt{\cos^2 x -\cos^2
    \chi_0}} \, \partial_\tau x 
,
\label{inhoN}
\end{eqnarray}
where $G_{\mu\nu}$ denotes again the target space metric and in the
second step we have used  (\ref{Ddefo}) to eliminate
$\tilde{\theta}$. The boundary term picked up in the $\theta$
variation of (\ref{GiKiDefo}) gives the same boundary condition as in
(\ref{inhoN}). Thus, we do not need to add $F$-flux to the boundary in
contrast to section \ref{recap}. (This confirms our earlier
statement that the model in\cite{Bachas:2000ik} and
\cite{Giveon:1993ph} are connected by a multi valued gauge
transformation.) In addition to (\ref{inhoN}) we find a Neumann
boundary condition for the second direction along the D2 brane,
\begin{eqnarray}
\lefteqn{\partial_\sigma x -\frac{R^2 \sin x\cos x \cos \chi_0}{\left( \cos^2
    x+ R^2 \sin^2 x\right) \sqrt{\cos^2 x -\cos^2 \chi_0}}\,
    \partial_\sigma \tilde{\theta} =}&&\nonumber \\ 
& & \;\;\;=\frac{ \sin x\cos x \cos 
    \chi_0}{\left( \cos^2 x+ R^2 \sin^2 x\right) \sqrt{\cos^2x -\cos^2
    \chi_0}}\, \partial_\tau \theta .
\end{eqnarray}

For consistency, we should check that the DBI action on
the D-brane is minimized by our boundary conditions. 
The DBI action reads (for vanishing $F$-fieldstrength)
\begin{equation}
S_{DBI} = T_{(2)}\int_{D2} d^2 \xi\, e^{-\Phi +\Phi_0}\sqrt{\det \hat{G}} ,
\label{DBIaction}
\end{equation}
where we identify $\Phi_0$ with the constant dilaton at $R=1$, and
$\hat{G}$ is the value of $G + B$ on the brane. The action
(\ref{DBIaction}) is invariant under reparameterizations of the
D-brane. We fix this invariance by the ``static'' gauge
\begin{equation}
x = \xi^1 \; ,\; \theta = \xi^2 \; ,\; \tilde{\theta}
=\tilde{\theta}\left( x\right) .
\end{equation}  
Here, we consider only boundary conditions preserving the
invariance under constant shifts of $\theta$.  
The metric
induced on the brane is 
\begin{equation}
ds_{D2}^2 =k\alpha^\prime \left( 1 + \frac{R^2 \cos^2 x \left(
      \partial_x\tilde{\theta}\right) ^2 }{ \cos^2
      x + R^2 \sin^2 x }\right)dx^2 + \frac{k\alpha^\prime\sin^2
      x}{\cos^2x + R^2 \sin^2 
      x} d\theta^2, 
\end{equation}
the induced $B$-field reads
\begin{equation}
B_{D2} = k\alpha^\prime \frac{\cos^2 x}{\cos^2 x +R^2 \sin^2 x} \partial_x
\tilde{\theta}\, dx d\theta ,
\end{equation}
and the dilaton is given by (see (\ref{dildefo}))
\begin{equation}
e^{\Phi_0 -\Phi} = \sqrt{\frac{\cos^2x + R^2 \sin^2 x}{R}} .
\end{equation}
Plugging this into (\ref{DBIaction}) yields
\begin{equation}
S_{DBI} = \frac{k\alpha^\prime}{R} T_{(2)}\int dx d\theta \sqrt{\sin^2
  x + \cos^2 x \left( \partial_x \tilde{\theta}\right)^2} .
\end{equation}
The corresponding equation of motion
\begin{equation}
\partial_x \left( \frac{\cos^2 x \partial_x
    \tilde{\theta}}{\sqrt{\sin^2 x +\cos^2 x \left( \partial_x
    \tilde{\theta}\right)^2}} \right) =0 
\end{equation}
is satisfied by our boundary condition (\ref{Ddefo}). Hence, we have
placed the D-brane in a conformal invariance preserving way. 

Multi valued gauge transformations (on $B$) can shift the $F$ flux
only by integer values and hence our result of vanishing $F$-flux
agrees with a quantization condition on the $F$-flux. To see this more
explicitly let us perform the following gauge transformation (first
in the closed string case)
\begin{equation}
\delta B = 2\pi \alpha^\prime d\Lambda ,
\label{Bgauge}
\end{equation}
with
\begin{equation}
\Lambda =\frac{k}{2\pi}\tilde{\theta}\, d\theta ,
\label{Lambda}
\end{equation}
being a multi valued function. 
The fact that the integral
\begin{equation}
\frac{1}{2\pi \alpha^\prime}\int \delta B = 2\pi k  
\end{equation}
is an integer multiple of $2\pi$ ensures that the wave function of a
fundamental 
string wrapping around $\delta B$ picks up an invisible Aharanov-Bohm
phase for integer $k$.
For the open string we have to combine (\ref{Bgauge}) with a shift
in the gauge fieldstrength taking us from the previously vanishing $F$
field to
\begin{equation}
F = - d\Lambda = -\frac{k}{2\pi}d\tilde{\theta}\wedge d\theta,    
\end{equation}
where (\ref{Ddefo}) should be imposed. 
The corresponding (integrated) $F$-flux is\footnote{Our computation
  assumes $\chi_0 \leq \frac{\pi}{2}$. For $\chi_0 \geq \frac{\pi}{2}$
  we just need to replace $\chi_0$ by $\pi -\chi_0$.}
\begin{equation}
\int_{D2} F = -k\int_{-\chi_0}^{\chi_0}d\tilde{\theta} = -2k\chi_0 ,
\label{intflux}
\end{equation}
which agrees with the flux obtained by integrating
(\ref{Fquant}). This derivation of the quantization condition for the
$F$-flux is completely independent of the value of the deformation
parameter $R$, and confirms our earlier statement that the position of
the D-brane should be given by (\ref{Ddefo}) for all $R$.
 
To summarize, we first discuss the D-branes for the model at
$R=\infty$. The $\theta$ direction decouples there and the D-brane
becomes a D1-brane on the remaining two dimensional surface. The range
for $x$ is restricted to $0\ldots\chi_0$ (for $\chi_0 \leq \frac{\pi}{2})$)
or $0 \ldots \pi -\chi_0$ (for $\chi_0 \geq \frac{\pi}{2}$). At
  $x =0$ the radius of the $\tilde{\theta}$ circle diverges. The
  boundary value for $\tilde{\theta}$ is given
  by $\cos\tilde{\theta}=\cos \chi_0$ there. When 
  $x$ takes its maximal size $\tilde{\theta}$ must be either $0$ or
  $\pi$ depending on the sign of $\cos\chi_0$. Thus the D-brane
  connects the points $x=0, \tilde{\theta} = \pm \chi_0$ along a
  geodesic through the two dimensional surface. 

\begin{figure}
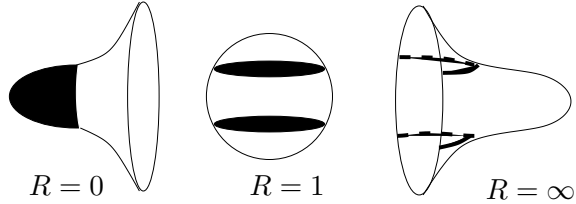
  
\begin{center}
\input branes.pstex_t
\end{center}
\caption{The ``middle'' and end points of the deformed model with the
  possible D-branes for $k=3$.}
\label{fig:branesdefo}
\end{figure}

At $R=0$ the $\tilde{\theta}$ direction decouples and the D2-brane
remains a two dimensional surface on the remaining geometry. The
maximal value for $x$ is again $\chi_0$ (or $\pi -\chi_0$). Hence, we
obtain the picture drawn in figure \ref{fig:branesdefo} (for $k=3$)
%%%%%%%%%%%%%%%%%%%%%%%%%%%%%%%%%%%%
%%%%%%%%%%%%%%%%%%%%%%%%%%%%%%%%%%%% 
Note, also that the D-branes fit into the picture that the two
endpoints of the deformation are related by T-duality. Performing a
T-duality with respect to $\tilde{\theta}$ of the $R=\infty$ model the
position of the D-brane becomes a gauge field component $A_\theta$ in
the $R=0$ model\cite{Alvarez:1996up},\cite{Dorn:1996an} (see also
section 2.3.3.2 of\cite{Forste:2001ah}).
Computing the corresponding fieldstrength one finds agreement with the
deformed model at $R=0$.

\section{Rotating branes and shifting Wilson lines}

In the class of models (parameterized by the deformation parameter
$R$), we are going to discuss the two following moduli. Firstly, we
can shift $\tilde{\theta}$ by a 
constant and secondly we can shift Wilson lines (constant values of
$A_\theta$). Shifting $\tilde{\theta}$ by a constant corresponds to
rotating the branes in the zero-three plane of the embedding ${\mathbb
  R}^4$ (see (\ref{depara})). Therefore, one could for example
construct configurations with more branes by adding to our previous
setup rotated branes. However, strings stretched between branes and
rotated branes will pull those branes on top of each other. Below, we
are going to argue that the shift symmetry in $\tilde{\theta}$ is
broken to ${\mathbb Z}_k$ for $R\to\infty$. This means that branes and
rotated branes cannot be pulled on top of each other in a continuous
way and hence we expect new stable configurations.

Our argument that the $U\!\left( 1\right)$ symmetries are broken to
${\mathbb Z}_k$ is very similar to the one given
in\cite{Maldacena:2001ky}. 
We consider the case with only closed strings.
Our deformed models can be viewed as a class of  $SU\!\left( 2\right)\times
U\!\left( 1\right) / 
U\!\left( 1\right)$ gauged models, where the deformation parameter
corresponds to the way the gauged $U\!\left( 1\right)$ is embedded in
$SU\!\left( 2\right) \times U\!\left( 1\right)$\cite{Sfetsos:1993ka}. 
To be explicit, we start with an $SU\!\left( 2\right) \times U\!\left(
  1\right)$ WZW model,
\begin{eqnarray}
S & = &  \frac{k}{2\pi}\int d^2 z \left\{ \partial_+ x\partial_- x +
  \sin^2 x \partial_+ \theta^\prime \partial_- \theta^\prime + \cos^2 x
  \partial_+ \tilde{\theta}\partial_- \tilde{\theta} \right. \nonumber
  \\
& & \left. \; +\cos^2 x\left( \partial_+ \theta \partial_-
  \tilde{\theta} -\partial_+ \tilde{\theta}\partial_- \theta\right)
  +\partial_+ y \partial_- y \right\} .
\label{product}
\end{eqnarray}
Now, we gauge a combination of constant shifts in $y$ and
$\theta^\prime$, promoting the corresponding symmetry to a local one.
This is done by replacing partial derivatives with the following
covariant derivatives ($\alpha$ is a worldsheet index)
\begin{eqnarray}
\partial_\alpha \theta^\prime &\rightarrow &\partial_\alpha\theta^\prime
+ \sqrt{R^2
  -1} A_\alpha , \\
\partial_\alpha y &\rightarrow &\partial_\alpha y+ A_\alpha
\end{eqnarray}
Under local shifts $\delta \theta^\prime$ the worldsheet gauge field
transforms as
\begin{equation}
\delta A_\alpha = -\frac{1}{\sqrt{R^2 -1}}\partial_\alpha \delta \theta^\prime
\end{equation}
and hence the size of the gauge group is $2\pi/ \sqrt{R^2 -1}$. The
action is invariant if we simultaneously shift $y$ by 
\begin{equation}
\delta y = \frac{1}{\sqrt{R^2 -1}}\delta \theta^\prime .
\end{equation}
This in turn fixes the size of the $U\!\left( 1\right)$ parameterized
by $y$. The gauge field $A_\alpha$ appears without derivatives in the
action and can be integrated out by replacing it with its classical
solution or solving the Gaussian integral. The resulting action is 
(\ref{GiKiDefo}) with $\theta = \theta^\prime -\sqrt{R^2 -1} y$. The
other combination $\theta^\prime + \sqrt{R^2 -1}y$ drops out. Its
integration cancels the integration over gauge equivalent fields
$A_\alpha$ up to a Jacobian which can be taken into account by
redefining the dilaton according to (\ref{dildefo}). Thus, for $R>1$ our
deformed model is equivalent to the gauged $SU\!\left( 2\right) \times
U\!\left( 1\right)$ model described above. Performing the multi valued
transformation (\ref{Bgauge}) with (\ref{Lambda}) (and $\theta$
replaced by $\theta^\prime$)\footnote{Alternatively one can switch on
  a constant $B_{y\tilde{\theta}}$ component, whose value is quantized
    by the requirement that the corresponding Aharanov-Bohm phase for
    the wave function of a string wrapped around this $B$-field is
    an integer multiple of $2\pi$. This will result in the same 
    quantization condition.} generates
  the following term in the
gauged $SU\!\left( 2\right) \times U\!\left( 1\right)$ model,
\begin{equation}
\delta S_{gauged} = \frac{k\sqrt{R^2 -1}}{2\pi} \int d^2 z\,
\tilde{\theta} F_{+-} ,
\end{equation}  
where $F_{+-}$ is the fieldstrength corresponding to
$A_\alpha$. Recalling that the size of the gauge group is
$2\pi/\sqrt{R^2 -1}$ one might conclude that the shift symmetry in
$\tilde{\theta}$ is broken to ${\mathbb Z}_k$ (i.e.\ only rotations
about integer multiples of $2\pi /k$ are allowed). 
However, performing another multi valued gauge transformation one can
also generate discrete values of a $B_{y\theta^\prime}$ component,
e.g.\
\begin{equation}
B = k\alpha^\prime \sqrt{R^2 -1}\, d\theta^\prime \wedge dy .
\end{equation}
After
gauging, the associated terms result for example in the following
coupling 
\begin{equation}
\delta S_{gauged} = \frac{k\sqrt{R^2 -1}}{2\pi} \int d^2 z \,
\theta F_{+-} .
\label{counterc}
\end{equation} 
Thus the global $U\!\left( 1\right) \times U\!\left( 1\right)$
transformations of shifting $\tilde{\theta}$ and $\theta$ are broken
to a $U\!\left( 1\right)\times {\mathbb Z}_k$. By shifting $B$ field
components by 
discrete amounts the embedding of the $U\!\left( 1\right)$ into the 
$U\!\left( 1\right)\times 
U\!\left( 1\right)$   can be changed. 
In the limit $R\to
\infty$ one gauges only the shift symmetry in $\theta$ and obtains the
quantization conditions on shifts in
$\tilde{\theta}$\cite{Maldacena:2001ky}.  
The connection to our discussion is obtained by imposing decoupling of
the $y$ direction for $R\to \infty$. This prohibits a $B_{y\theta}$
term and hence the coupling (\ref{counterc}) cannot be generated.  
One could be tempted to conclude
that a similar argument provides a quantization condition in the other
limit $R=1$. In this limit one would gauge only shifts in the $y$
direction and thus the size of the additional $U\!\left( 1\right)$
would be a modulus. Hence, no quantization condition on the constant
$\tilde{\theta}$ shifts arises in the undeformed model. 

By gauging a suitable combination of shifts in $\tilde{\theta}$ and
shifts in $y$ in (\ref{product}) one can also relate the models with
$R<1$ to gauged $SU\!\left( 2\right) \times U\!\left( 1\right)$
models. At $R=0$ one finds that shifts in $\theta$ are broken to
a discrete ${\mathbb Z}_k$ symmetry, if one imposes decoupling of the
$y$ direction in the limit $R\to 0$. Wilson lines $C_\theta$ can be 
written as 
\begin{equation}
iC_\theta  =  e^{-iC_\theta\theta}\partial_\theta
e^{iC_\theta\theta}
\label{wilsonloop}
\end{equation}
where $\theta$ is the center of mass position of the open string
ending on the $D2$ brane, and $C_\theta$ is a pure gauge field only
for integer $C_\theta$, because we have to require periodicity under
$2\pi$ shifts in $\theta$ for gauge transformations. Hence, there is a
one-to-one correspondence between the moduli space of Wilson lines and
the space of center of mass positions of the open string. If the
space of center of mass positions of the open string is broken from
$U\!\left( 1\right)$ to ${\mathbb Z}_k$ the definition
(\ref{wilsonloop}) does not make 
sense because the $\theta$ derivative is ill defined. We assume that
the moduli space of Wilson lines is broken to ${\mathbb Z}_k$. This
assumption is supported by the requirement that the two endpoints of
the deformation line are related by T-duality and for $R=\infty$ the
branes can be rotated only by an integer times $\frac{2\pi}{k}$.

\section{The A- and B-branes}

So far, we have considered only the deformation of branes leaving an
$SU\!\left( 2\right)$ invariant in the undeformed model. These are the
branes wrapped on ``parallel'' $S^2$ submanifolds of $S^3$, i.e.\ the
$S^2$'s have an identical rotation axis. In such a setup the $SO\!\left(
  4\right)$ (with universal covering $SU\!\left( 2\right)\times
SU\!\left( 2\right)$) isometry is broken to an $SO\!\left( 3\right)$
(with universal covering $SU\!\left( 2\right)$) subgroup. If we rotate
the rotation axis of one brane with respect to the other, each brane
leaves a different subgroup of $SO\!\left( 4\right)$ invariant and
hence the isometry is completely broken. In the undeformed case such a
setup should not be stable, since open strings stretching between the
two branes will `pull' them to the symmetric setup. However, for
$R\to\infty$ the $U\!\left( 1\right)$ rotation group is broken to
${\mathbb Z}_k$ and hence there can be new stable setups. These setups
we are going to discuss in the following.

\begin{figure}
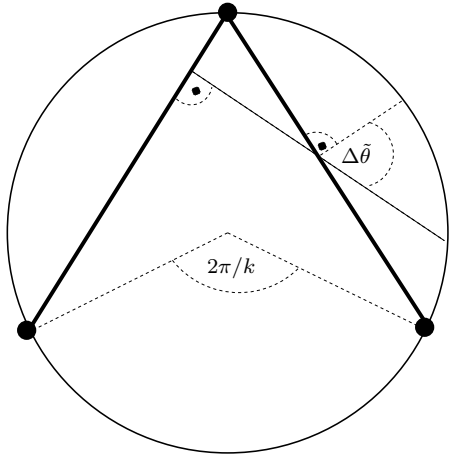
  
\begin{center}
\input abranes.pstex_t
\end{center}
\caption{A typical A-brane at $R=\infty$ for $k=3$.}
\label{fig:abranes}
\end{figure}

Let us restrict to the case $R>1$. (For $R<1$ relative angles between
branes should be replaced by relative angles between Wilson lines.) We
start with two $SU\!\left( 2\right)$ preserving D2-branes (at
coinciding or different locations). Then we rotate one of the D2
branes untill they touch each other in one point and form a new D2
brane. For $R\to\infty$ we cannot rotate the branes back into a
``parallel'' position in 
a continuous way. At that point the $S^3$ transforms into a
geometry which is topologically a disc times a circle. On the disc our
brane setup takes the form depicted in figure \ref{fig:abranes}. This 
corresponds to the A-branes discussed in\cite{Maldacena:2001ky}.

%%%%%%%%%%%%%%%%%%%%%%%%%%%%%%%%%%%%
%%%%%%%%%%%%%%%%%%%%%%%%%%%%%%%%%%%% 

In the following we are going to investigate the A-branes 
along the line $R>1$. Let us restrict to the setup
obtained from two D2-branes with originally coinciding
positions. After the rotation the position of the second brane becomes
\begin{equation}
\cos \chi_0 = \cos x \cos \left(\tilde{\theta} + \Delta\tilde{\theta}\right) ,
\end{equation}
where 
\begin{equation}
\Delta\tilde{\theta} =\frac{\pi}{k} .
\end{equation}
For $R>1$ we have D2-branes. These
should satisfy the $F$-flux quantization condition.
The total $F$-flux of the A-brane is computed to\footnote{Again, we
  assume $\chi_0 \leq \frac{\pi}{2}$. If this is not the case we
  replace $\chi_0$ by $\pi -\chi_0$ in the result.}
\begin{equation}
\int_{D2} F = -k \int_{-\chi_0}^{\chi_0} d\tilde{\theta} -k\int_{-\chi_0
  +\Delta\tilde{\theta}}^{\chi_0 -\Delta\tilde{\theta}}d\tilde{\theta}
  = -2k\left( 
  2\chi_0 -\Delta\tilde{\theta}\right) \equiv -2k\chi_0 ^\prime ,
\label{afflux}
\end{equation}
where in the last step we have written the $F$-flux in the form of
(\ref{intflux}). Since $\chi_0^\prime = 2\chi_0 -\pi/k$ satisfies the
quantization condition (\ref{quantization}) the considered A-brane
satisfies the $F$-flux quantization condition. 
For our particular example, the $F$-flux coincides with the $F$-flux
of a single D2-brane located at $\chi_0$. This is consistent with the
statement in\cite{Maldacena:2001ky} that the `elementary' A-branes
consist out of straight lines connecting pairs of the $k$
points on the disc and $D0$ branes located at the ${\mathbb Z}_k$
points. (Since for a single brane in $SU\!\left( 2\right)$ the
rotation is irrelevant, we can identify each deformed single $SU\!\left(
  2\right)$ brane with a straight line connecting two of the $k$
points. However, a set of $SU\!\left( 2\right)$ preserving branes with
different locations leads to parallel lines on the disc.)
Extending our analysis
to more general A-branes in a straightforward way one can show that
all possible A-branes satisfy the $F$-flux quantization condition.

One can also start with more than two D2-branes and rotate the second
brane such that it touches the first one in a point and the third
brane such that it touches the second one in a point and so on. Then
the value of $\chi_0^\prime$ in (\ref{afflux}) becomes
\begin{equation}
\chi_0^\prime = \sum_{i=1}^N \chi_i -\sum_{i=1}^{N-1} \Delta 
\tilde{\theta}_i 
\label{shiftedchi}
\end{equation}  
where $\chi_i$ are the original positions of the branes and
$\Delta\tilde{\theta}_i$ is the relative angle between the first and
the $i+1$'st brane. 

\begin{figure}
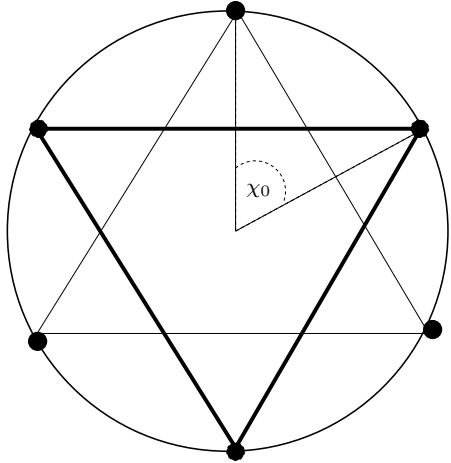
  
\begin{center}
\input bbranes.pstex_t
\end{center}
\caption{A typical B-brane setup at $R=\infty$ for $k=6$. The triangle
  drawn in softer lines corresponds to $\chi_0 \to \pi
  -\chi_0$. Adding this configuration leads to a ${\mathbb Z}_k$
  invariant picture.}
\label{fig:bbranes}
\end{figure}

Now, let us consider the following configuration at $R=1$. We start
(for even $k$)
with $k/2N$ D2-branes located at $\chi_0 = \frac{2N\pi}{k}$, where
$N$ is a positive integer such that $k/2N$ is an integer.  
Next, we rotate the second brane by an angle $\frac{2N\pi}{k}$, the
third brane by $\frac{4N\pi}{k}$ and so on. In the $R\to \infty$ limit
this yields the setup drawn in figure \ref{fig:bbranes} (for $k=6$).
%%%%%%%%%%%%%%%%%%%%%%%%%%%%%%%%%%%%
%%%%%%%%%%%%%%%%%%%%%%%%%%%%%%%%%%%% 
This is the T-dual version of the B-branes discussed
in\cite{Maldacena:2001ky}. At $R=\infty$ it is not stabilized by an
$F$-flux, since (after T-dualizing the decoupled $\theta$ direction)
the D2-branes become D1-branes which do not support an
$F$-flux. Another way, to stabilize branes could be given by the
${\mathbb Z}_k$ quantization of shifts in $\tilde{\theta}$. However,
the B-branes form closed loops and can decay by shrinking to zero
branes at the center of the disc\cite{Maldacena:2001ky}. (In the
T-dual description the B-branes correspond to D2 branes covering only
a part of the disc. Those D2 branes can collapse to zero
branes.)  For the B-brane setup at finite $R$ we learn from
(\ref{shiftedchi}) that the $F$-flux generically vanishes (modulo
$2\pi k$). Therefore, 
we expect that the B-branes are not protected by an $F$-flux
quantization condition. Since they form closed loops also for finite
$R\geq 1$ they can collapse in a ${\mathbb Z}_k$ invariant way also
along the deformation line. An exception is given for $\chi_0
=\frac{\pi}{2}$. In this case a ${\mathbb Z}_k$ invariant setup is
formed by $k/2$ branes connecting antipodal points with relative angles
$2\pi /k$. These do not form closed loops
and are protected from decay by the ${\mathbb Z}_k$ quantization
condition of shifts in $\tilde{\theta}$. At $R<\infty$ this quantization
condition is lifted and the B-branes can continuously rotate. This
will change the $F$-flux and the system becomes unstable. (In the
conformal field theory prescription this should correspond to a
relevant boundary perturbation. Such perturbations are studied
in\cite{Recknagel:2000ri,Fredenhagen:2001nc}.)
The $F$-flux is preserved if the number 
of D2 branes decreases and D0 branes might be created. 
(Since the D0 branes do not carry $F$-flux we cannot determine their
existence within our present approach. A description of spherical D2
branes in terms of (noncommuting) D0 branes is given
in\cite{Myers:1999ps}, see also \cite{Fredenhagen:2001nc} for the
$SU\!\left(2\right)$ WZW model. In particular, the creation of D0
branes can be attributed to a conservation of brane charges taking
values in $K^{*}_H\!\left( SU\!\left( 2\right)\right) = {\mathbb
  Z}_k$\cite{Bouwknegt:2000qt,Fredenhagen:2001nc}.) 
Deformed back to the $R=\infty$ 
point the rotated B-brane will be seen as a ${\mathbb Z}_k$
symmetric setup with less D1 branes stretching between antipodal
points on the disc, and the missing D1 branes are replaced by D0
branes.  
Since the $F$-flux of the ${\mathbb Z}_k$ symmetric setup vanishes one
can obtain a system with D0 branes, only. The description of
B-branes in terms of D0-branes is given in\cite{Maldacena:2001ky}. 
Our view on the phase transition of B-branes is similar to the
way phase transitions among stable non BPS D-branes can be connected
to marginal deformations (for reviews see e.g.\ \cite{Sen:1999mg},
\cite{Gaberdiel:2000jr}, \cite{Lerda:1999um}).

\section{Orientifold fixed planes}

\begin{figure}
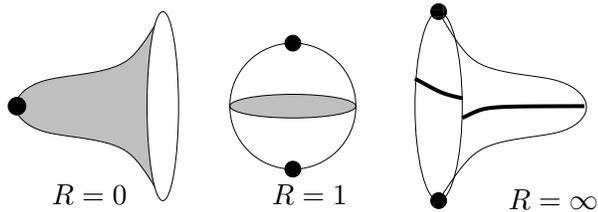
  
\begin{center}
\input planes.pstex_t
\end{center}
\caption{The ``middle'' and end points of the deformed model with the
  possible O-planes.}
\label{fig:planes}
\end{figure}

In this section we will briefly describe the location of orientifold
fixed planes in the models along the line of marginal
deformations. Early investigations of orientifolds of the $SU\!\left(
  2\right)$ models can be found in\cite{Johnson:1997ib},
\cite{Forste:1997ur}, \cite{Bianchi:1997gt}. More recent thorough
studies are performed in\cite{Brunner:ft}, \cite{Huiszoon:2001wv},
\cite{Bachas:2001id}. Here, we just give the fixed point sets under
symmetries containing a worldsheet parity reversal, i.e.\ the O-planes.
In the WZW model worldsheet parity transformations leave the action
invariant provided that we combine them with replacing the group
element $g$ by $\pm g^{-1}$. For the plus sign this gives the
antipodal points $g=\pm 1$ as fixed planes whereas for the minus sign
the equatorial $S^2$ is left invariant. If we parameterize the group
element as in (\ref{depara}) the possible orientifold transformations
read
\begin{equation}
x\left( z_+ , z_-\right) \to -x\left( z_- , z_+\right) \; ,\;
\tilde{\theta}\left( z_+ , z_-\right) \to -\tilde{\theta}\left( z_- ,
  z_+\right) \; ,\; \theta\left( z_+, z_-\right) \to \theta\left( z_-
  , z_+\right) ,
\end{equation}
and   
\begin{equation}
x\left( z_+ , z_-\right) \to x\left( z_- , z_+\right) \; ,\;
\tilde{\theta}\left( z_+ , z_-\right) \to \pi -\tilde{\theta}\left( z_- ,
  z_+\right) \; ,\; \theta\left( z_+, z_-\right) \to \theta\left( z_-
  , z_+\right) .
\end{equation}
It is easy to see that also the deformed model (\ref{GiKiDefo}) is
invariant under these orientifold transformations. Hence the
corresponding O-planes  can be followed
along the deformation line. Figure \ref{fig:planes} shows the possible
O-planes in 
the undeformed model and for the two endpoints of the deformation
line.
%%%%%%%%%%%%%%%%%%%%%%%%%%%%%%%%%%%%
%%%%%%%%%%%%%%%%%%%%%%%%%%%%%%%%%%%% 
Again the models at the end of the deformation line are related by
T-duality if we view the O-zero-plane at the left tip of the left
picture as a degenerated O-two-plane.
%%%%%%%%%%%%%%%%%%%%%%%%%%%%%%%%%%%%%%%%%%%%%%%%

\section{Conclusions}

In the present article we gave a semiclassical description for
D-branes along a line of marginally deformed $SU\!\left( 2\right)$ WZW
models. At the ends of the line one obtains $SU\!\left(
  2\right)/U\!\left( 1\right)$ coset models.
At the $SU\!\left( 2\right)$ point the moduli space of rotating
branes and Wilson lines is $U\!\left( 1\right)\times U\!\left(
  1\right)$ whereas it is 
broken to $U\!\left( 1\right)\times {\mathbb Z}_k$ away from the
$SU\!\left( 2\right)$ 
point. This suggests a mechanism for phase transitions in brane
configurations of coset models. One can marginally deform the coset
configuration to the $SU\!\left( 2\right)$ point where phase
transitions can be triggered by rotating the branes such that the
$F$-flux quantization condition is violated. Since the $F$-flux
quantization is not directly connected to conformal invariance we are
not sure whether the phase transition can be associated to a relevant
perturbation of the CFT description. Assuming that this is the case,
we can use known results\cite{Fredenhagen:2001nc} to identify the
phase transition as a transition between D2 and D0 branes. The
geometric picture suggested by our analysis is, that at the
$SU\!\left( 2\right)$ point D2 branes at
angles transform into less D2 branes and D0 branes. (The term `at
angles' refers to an angle between the rotation axes of the D2 branes
through the center of $S^3$.)
The transformed setup can be marginally deformed to the coset point
where it can be associated to a phase transition in the coset model.

It should be interesting to extend our discussion to a more general
analysis in deformed WZW models and beyond the semiclassical level. A
CFT treatment is the subject of ongoing work\cite{daniel} and
preliminary results agree with the presented discussions.
Another open question is to give an effective field theory description
of the presented models. Using the observation that each point along
the line of deformations can be identified with an $SU\!\left(
  2\right) \times U\!\left( 1\right) /U\!\left( 1\right)$ coset this
should be a straightforward application of the results presented
in\cite{Fredenhagen:2001kw}. 
New features might arise for non compact groups. In non compact
groups the Cartan-Killing metric is indefinite and therefore
qualitatively different deformations are possible (into space-like,
time-like and null-directions)\cite{Forste:1994wp}.
  
\vskip 1cm  
  
\noindent {\bf Acknowledgments} 
 
\noindent 
I would like to thank Daniel Roggenkamp for very useful discussions.
This work is supported by the European Commission RTN programs
\mbox{HPRN-CT-2000-00131}, 00148 and 00152.

\end{document}